\documentclass[lettersize,journal]{IEEEtran}
\usepackage{amsmath,amsfonts}
\usepackage{algorithmic}
\usepackage{algorithm}
\usepackage{array}
\usepackage[caption=false,font=normalsize,labelfont=sf,textfont=sf]{subfig}
\usepackage{textcomp}
\usepackage{stfloats}
\usepackage{url}
\usepackage{verbatim}
\usepackage{graphicx}
\usepackage{cite}
\usepackage{booktabs}
\begin{document}

\title{When Wires Can’t Keep Up: Reconfigurable AI Data Centers Empowered by Terahertz Wireless Communications}

\author{Chong Han, Mingjie Zhu, Wenqi Zhao, Ziming Yu, Guolong Huang, Guangjian Wang, Wen Tong, Wenjun Zhang}




\maketitle

\begin{abstract}
The explosive growth of artificial intelligence (AI) workloads in modern data centers demands a radical transformation of interconnect architectures. Traditional copper and optical wiring face fundamental challenges in latency, power consumption, and rigidity, constraining the scalability of distributed AI clusters. This article introduces a vision for Terahertz (THz) Wireless Data Center (THz-WDC) that combines ultra-broadband capacity, one-hop low-latency communication, and energy efficiency in the short-to-medium range (1–100m). Performance and technical requirements are first articulated, including up to 1 Tbps per link, aggregate throughput up to 10 Tbps via spatial multiplexing, sub-50 ns single-hop latency, and sub-10 pJ/bit energy efficiency over 20m. To achieve these ambitious goals, key enabling technologies are explored, including digital-twin-based orchestration, low-complexity beam manipulation technologies, all-silicon THz transceivers, and low-complexity analog baseband architectures. Moreover, as future data centers shift toward quantum and chiplet-based modular architectures, THz wireless links provide a flexible mechanism for interconnecting, testing, and reconfiguring these modules. Finally, numerical analysis is presented on the latency and power regimes of THz versus optical and copper interconnects, identifying the specific distance and throughput domains where THz links can surpass conventional wired solutions. The article concludes with a roadmap toward wireless-defined, reconfigurable, and sustainable AI data centers.
\end{abstract}


\section{Introduction}
\IEEEPARstart{O}{ver} the past decade, the exponential increase in AI model complexity, e.g., large-scale deep learning and generative models, has pushed data-center infrastructure to its limits~\cite{8367741}. Training a trillion-parameter model may require tens of thousands of GPUs or AI accelerators interconnected through high-bandwidth, low-latency fabrics. Today’s copper and optical interconnects, although mature, suffer from limitations in flexibility, latency, and energy efficiency. Copper traces encounter high attenuation and skin-effect losses above tens of gigahertz, necessitating complex equalization and consuming considerable power. Optical fibers, while capable of high data rates, introduce energy overhead through electro-optical conversion and static transceiver power consumption. Moreover, wired links enforce fixed topologies, making it difficult to adapt to dynamic computational workloads and traffic patterns inherent in AI clusters.

\begin{figure*}
    \centering
    \includegraphics[width=1.0\linewidth]{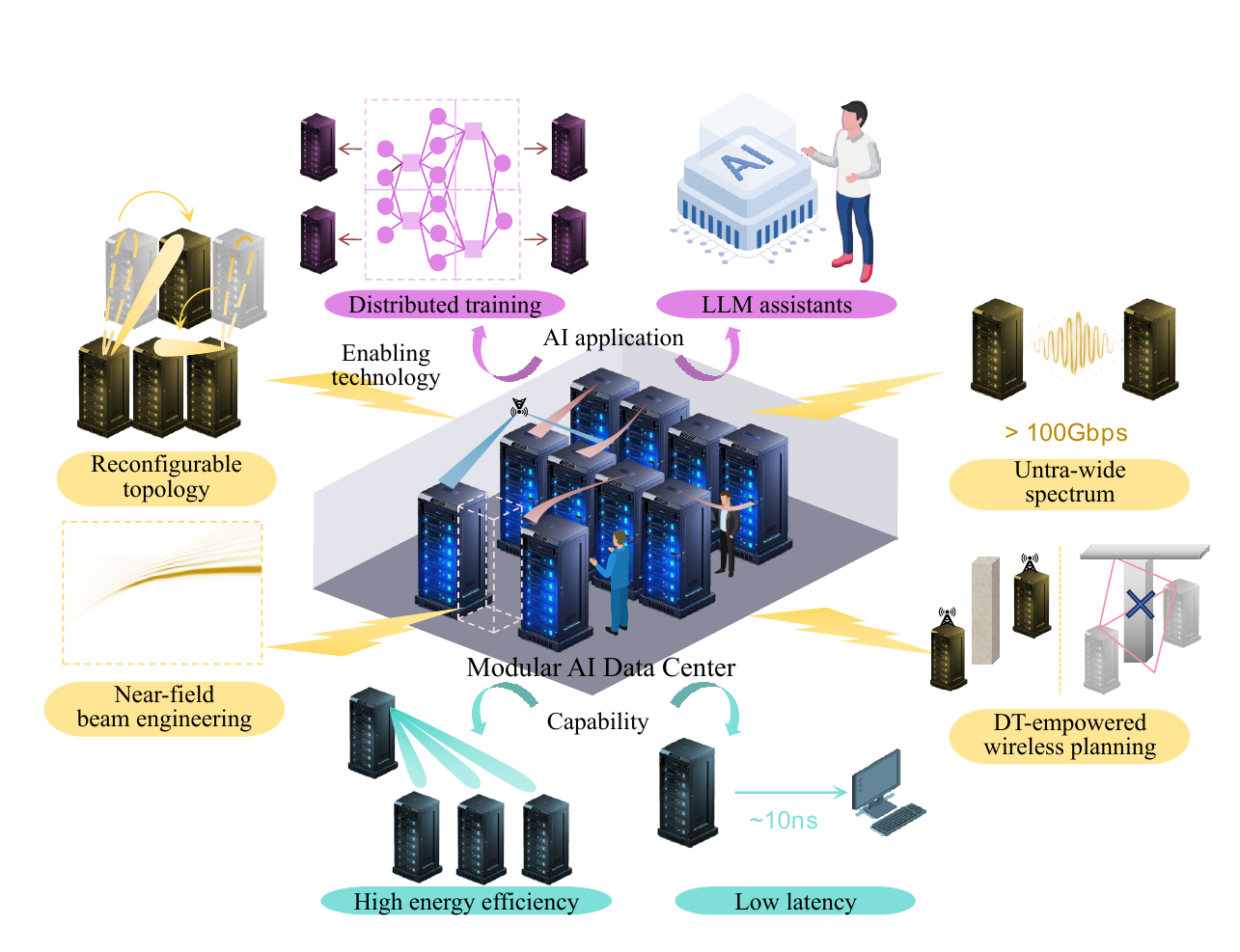}
    \caption{A conceptual architecture of THz wireless interconnects in an AI cluster.}
    \label{fig:ConArc}
\end{figure*}

The vision of Terahertz (THz) Wireless Data Center (THz-WDC) seeks to overcome these bottlenecks by replacing or complementing wired connections with short-range, highly directional THz wireless links~\cite{7393451}. Operating in the 0.1–1 THz spectrum, these links can provide ultra-high bandwidth and nanosecond-scale latency while eliminating the inflexibility and maintenance challenges of physical cables~\cite{6005345, 9766110}. In essence, the THz-WDC aims to enable an intelligent, reconfigurable compute fabric where connectivity dynamically 
adapts to task demands, a paradigm we term “\textit{shape follows task}”.
Looking ahead, future data centers are expected to adopt modular architectures built around quantum computing units~\cite{doi:10.1126/science.adz8659}, and THz wireless links offer the most straightforward and non-intrusive means of interconnecting and testing these modules.
Fig.~\ref{fig:ConArc} illustrates the conceptual architecture of THz wireless interconnects within an AI computing cluster.
First, the direct single-hop point-to-point THz links eliminate the need for multi-stage electrical or optical switching, thereby providing ultra-low hop latency and reducing the end-to-end communication delay across accelerated nodes.
Second, the dynamically reconfigurable topology enabled by electronically steerable THz beamforming supports self-adaptive link management. This capability allows the network to rapidly allocate bandwidth, reroute traffic, and recover from blockages, significantly improving overall network utilization and efficiency in large-scale distributed training workloads.
Third, besides using Gaussian beams as classically considered in beamforming and beamfocusing, the use of airy beams enables robust quasi-LoS connectivity by generating self-healing and self-bending propagation paths that can bypass partial obstructions and maintain stable link quality even in dense rack environments.
Moreover, by integrating a digital twin (DT) of the AI cluster, the system can continuously monitor link states, predict interference patterns, and proactively optimize routing and beam configurations. This closes the loop between physical-layer measurements and high-level orchestration, enabling globally optimized THz interconnect performance.

In this article, we first propose the architectural wireless AI cluster, then give the system performance targets the THz wireless AI cluster can achieve. The enabling technologies, including DT, beam manipulation technologies, THz tranceivers, and baseband processing, are introduced, followed by a comparison of latency and energy efficiency with traditional copper or optical fiber. Finally, challenges and future directions are elaborated on this promising paradigm.

\section{Architectural Vision: Wireless-Enabled AI Compute Clusters}
Future AI workloads will not only demand massive computational throughput but also flexible communication topologies, hence a low-latency and high-energy-efficiency inter-communication is a need to make sure the explosive data stream can be processed and transmitted in time. A THz wireless data center allows dynamic formation of compute clusters based on current job requirements. Racks or boards can establish direct one-hop THz links through beam steering, bypassing multiple switch and router stages. This flexibility enables reconfigurable topologies such as 2D meshes, 3D tori, or ad-hoc all-to-all links, formed on demand via electronic beam reconfiguration. Therefore, the extreme delay from switches would be eliminated, and the latency would be obviously decreased.
For distributed AI training, where gradient exchange and model synchronization dominate traffic, such adaptive connectivity can dramatically reduce queueing and serialization delays. Moreover, with multi- transmission and broadcast capability, the interconnect can efficiently handle collective operations such as AllReduce, a distributed computing primitive in which all nodes aggregate (e.g., sum) their local tensors and then share the global result with each other, thereby accelerating synchronization-heavy workloads like gradient updates in large-scale AI training.

\begin{figure*}[t]
    \centering
    \includegraphics[width=0.7\linewidth]{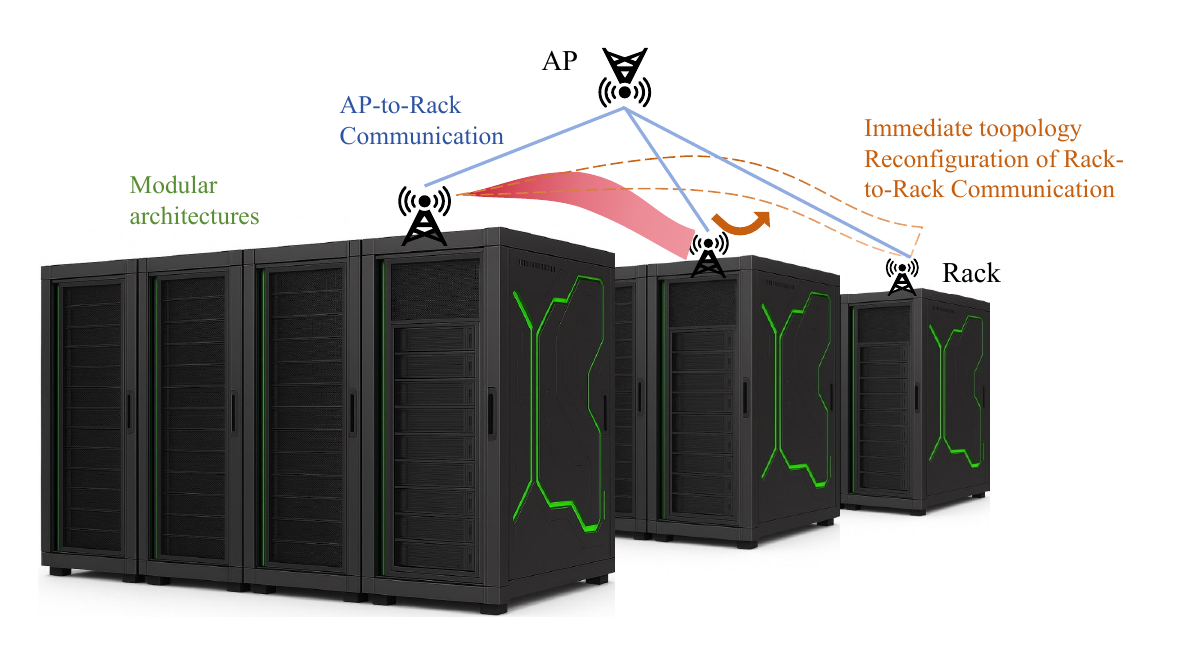}
    \caption{Dynamic reconfigurable topology enabled by THz links. }
    \label{fig:reconTHzToop}
\end{figure*}

Fig.~\ref{fig:reconTHzToop} illustrates the concept of a dynamically reconfigurable topology enabled by THz wireless links within an AI-driven computing cluster. By leveraging steerable high-capacity THz beams, the network can adapt its inter-rack connectivity in real time to match the workload demands. In the figure, an AI-driven orchestrator monitors traffic patterns and selectively activates high-bandwidth THz links while deactivating or downgrading others via beam manipulation. This flexibility allows the cluster to transition between different topologies, such as bandwidth-optimized, latency-optimized, or load-balanced configurations, without any physical rewiring. Such adaptive reconfiguration represents a fundamental shift from static wired networks toward wireless-defined, workload-aware interconnect fabrics.

\section{System Performance Requirements}


Based on emerging hardware research and modeling, the system-level targets of a THz-WDC can be summarized in Table~\ref{target}. These targets represent the performance envelope required for future AI-scale data centers, where cluster-wide training jobs demand increasingly tight coupling between compute nodes.
To contextualize the table, each metric highlights a fundamental limitation of conventional wired infrastructures and a corresponding opportunity for THz wireless links. First, the extremely wide contiguous bandwidth available in the THz band enables multi-band aggregation and dual-polarization signaling, bringing per-link peak rates to the terabit scale, which far exceeds what a single optical fiber or copper cable can provide without deploying multiple parallel lanes. Second, unlike multi-hop optical switching fabrics that introduce microsecond-level delays through queuing, switching, and serialization, a THz wireless hop enables near line-of-sight (LoS), single-hop communication with propagation-limited latency on the order of tens of nanoseconds.

These metrics capture the fundamental advantages that short-range, high-directionality THz links can offer, namely multi-terabit data rates, ultra-low hop latency, improved energy efficiency, and unprecedented reconfigurability. Together, they outline the design envelope that future THz-based data center architectures aim to achieve.
Energy efficiency is another critical dimension. While optical interconnects require high-power drivers, modulators, and thermal stabilization, and copper traces suffer from skin-effect losses at high bandwidth, THz transceivers based on tightly integrated CMOS/SiGe front-ends have the potential to operate at sub-10 pJ/bit over short to medium ranges (10–100 m). 
This opens the door to significantly reducing the overall network power budget.
Furthermore, THz links enable dynamic topology reconfiguration, allowing the interconnect fabric to adapt to evolving AI job structures, traffic hotspots, and resource scheduling demands. Such flexibility is fundamentally difficult to achieve with static cabling in today's data centers and motivates the search for wireless alternatives with controllable beam patterns and digitally orchestrated connectivity.

These requirements are aspirational but grounded in realistic trends observed in THz transceiver research, on-chip wireless interconnect demonstrations, and beamforming efficiency improvements. Recent prototypes in the 280–320 GHz band have shown multi-GHz bandwidths~\cite{11048931}, over 20 dBi integrated antenna gain, and improved LO phase noise, all of which suggest a plausible roadmap toward practical THz-WDC deployments. Notably, the latency and power metrics indicate potential orders-of-magnitude gains if THz front-ends achieve compact, low-power operation through deep co-integration of RF, antenna arrays, and mixed-signal baseband in advanced CMOS and SiGe processes.
\begin{table*}[!ht]
    \centering
    \caption{The system-level targets of a THz-WDC}
    \renewcommand\arraystretch{1.5}
    \makebox[\textwidth][c]{%
    \begin{tabular}{|c|c|c|}
    \hline
        \textbf{Metric} & \textbf{THz Wireless Target} & \textbf{Current Optical/Copper Baseline} \\ \hline
        Peak data rate (single link) & $0.1-1$ Tbps (dual-pol, multi-band) & $400–800$ Gbps  \\ \hline
        Aggregate throughput & $5–10$ Tbps (via MIMO/spatial aggregation) & $\leq 3$ Tbps (multi-fiber)  \\ \hline
        Link latency & \textless $50$ ns (one-hop direct) & $1-5$ µs (multi-hop optical fabrics) ~\cite{7828146} \\ \hline
        Energy efficiency (without baseband signal processing) & \textless $10$ pJ/bit \textless 20~m; \textless $1$~pJ/bit \textless 5~m  & $20–40$ pJ/bit (optical); $50–100$ pJ/bit (copper)  \\ \hline
        Topology flexibility & Dynamic beam-formed links & Static cabling \\ \hline
    \end{tabular}
    }
    \label{target}
\end{table*}

\section{Enabling Technologies}
\begin{figure*}
    \centering
    \includegraphics[width=1.0\linewidth]{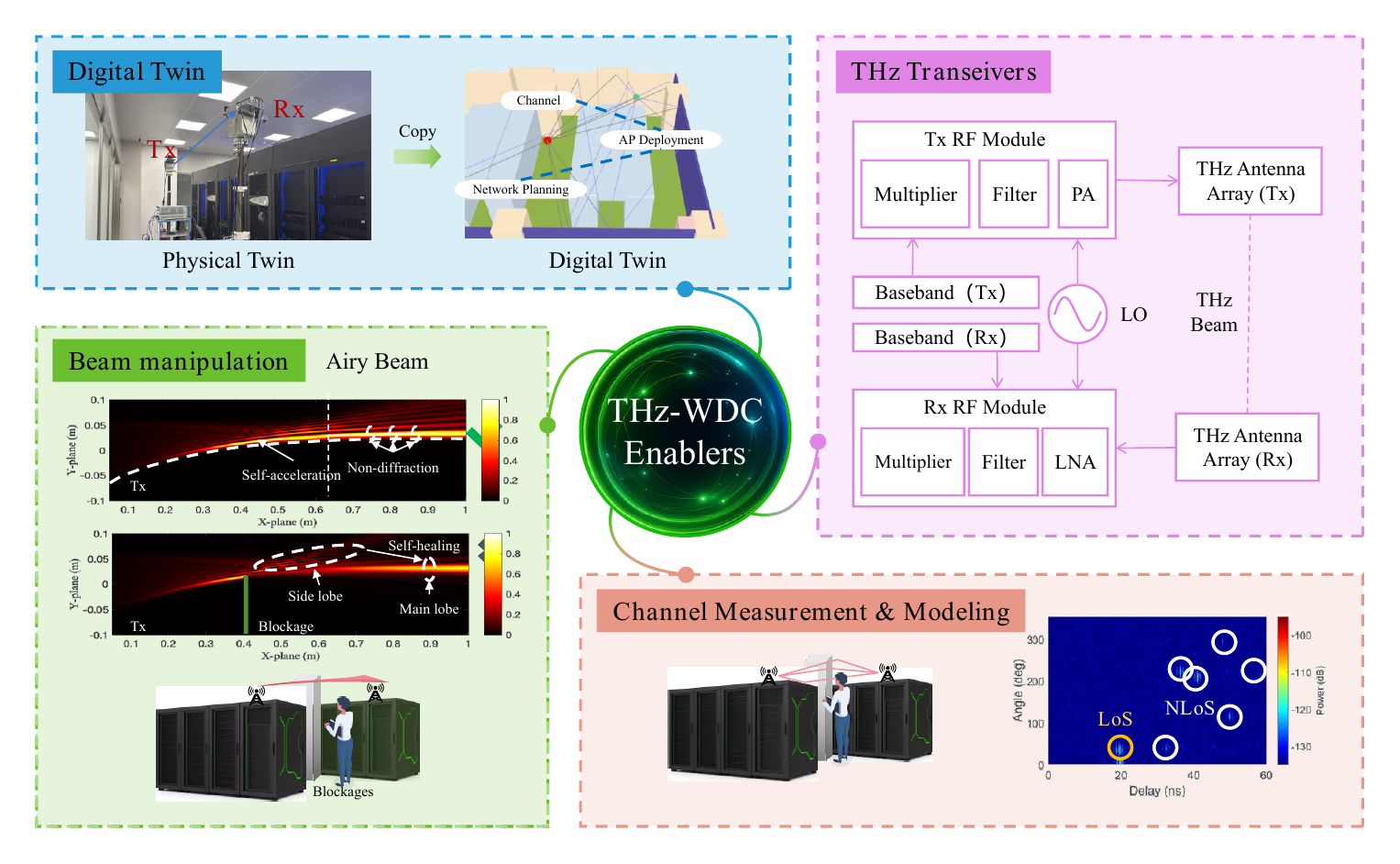}
    \caption{Enabling technolgoies for THz WDC, including digital-twin-based orchestration,  manipulation, silicon-based THz transceivers, and ultra-low-power analog baseband architectures.}
    \label{fig:thzenabler}
\end{figure*}

Several technology pillars underpin the THz-WDC vision. These include digital-twin-based orchestration, beam  manipulation, silicon-based THz transceivers, and ultra-low-power analog baseband architectures, illustrated in Fig.~\ref{fig:thzenabler} and detailed as follows.

\begin{itemize}
    \item \textbf{Digital twin-driven network orchestration}: By continuously sensing the THz channel conditions through integrated sensing-communication nodes, the network can maintain a real-time DT of the data center. This twin enables AI-assisted beam scheduling, interference management, and task-to-link matching, ensuring optimal throughput and reliability.
    Through DT technology~\cite{11039666}, efficient AP deployment and real-time network planning and management can be achieved.
    DT is supported by environment reconstruction, channel measurement with deterministic modeling, and backend optimization strategies. Environment reconstruction relies on multimodal fusion of infrared scanning and visible-light imaging, together with material properties obtained from THz-TDS measurements. Channel measurements characterize both LoS and NLoS propagation paths, providing alternative solutions when the LoS route is unavailable. Meanwhile, optimization methods targeting system capacity guide AP deployment and facilitate real-time resource allocation.

    \item \textbf{Measurement-based hybrid channel modeling} Accurate channel knowledge is fundamental for designing reliable THz wireless interconnects in data centers, where dense metallic structures and narrow aisles produce rich multipath and frequent obstruction. Measurement-driven analysis~\cite{10609428} enables identification of dominant paths, reflection surfaces, and power distribution across the environment—insights that directly guide transceiver placement,  management policies, and link scheduling strategies. To capture these characteristics, we conduct THz channel measurements using a VNA-based channel sounder operating from $290$ to $310$ GHz. A directional scanning scheme at the receiver samples the azimuthal plane from $0^\circ$ to $355^\circ$ and the zenith from $-20^\circ$ to $20^\circ$ in $5^\circ$ steps. From the resulting power–angle–delay profiles (PADPs), dominant multipath components can be extracted and matched against ray-tracing predictions, enabling hybrid data-driven and deterministic channel modeling. Path-loss trends at different distances can be directly characterized, as illustrated in Fig.~\ref{fig:scenario}(b), providing empirical validation for large-scale models. Furthermore, the measurements reveal that only about 52\% of links maintain an obstructed line-of-sight (OLoS), underscoring the importance of robust steering, non-line-of-sight (NLoS) exploitation, and topology-aware scheduling in practical data-center deployments.

    \begin{figure}
        \centering
        \subfloat[]{
            \includegraphics[width=0.85\linewidth]{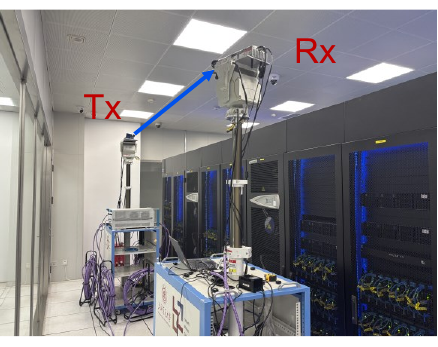}
        }
        \hfill
        \subfloat[]{
            \includegraphics[width=0.9\linewidth]{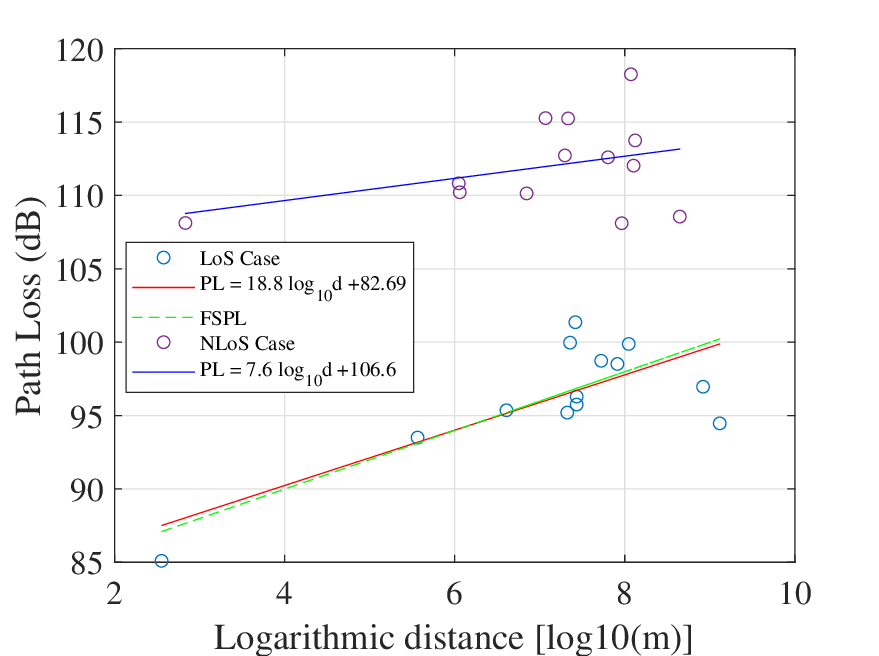}
        }
        \caption{Channel measurement for an AI cluster. (a) Measurement scenario. (b) Path loss modeling.}
        \label{fig:scenario}
    \end{figure}

    \item \textbf{Low-complexity near-field beam manipulation technologies}: To overcome high free-space path loss, directional beams are necessary. Hybrid architectures combining analog phase arrays, dielectric lenses, and RIS (Reconfigurable Intelligent Surfaces) can deliver multi-beam operation with low hardware complexity and energy overhead.
    Within the WDC, near-field beam technologies such as airy beam, which exhibits non-diffraction, self-healing behavior, and naturally accelerates along a curved propagation trajectory, are a promising technique to mitigate quasi-LoS propagation and enhance spectral efficiency in THz wireless data centers~\cite{zwq2}. Meanwhile, the multi-user widely-spaced array~\cite{10902604} enabled by fluid antenna systems can be applied to enlarge the near-field region to introduce the additional distance domain, and lead to a new paradigm of cross-near-and-far-field communications. 


    \item \textbf{Advanced THz transceivers}: Recent progress in electronic device technologies, particularly CMOS and SiGe BiCMOS, has enabled oscillation frequencies approaching 380 GHz~\cite{10870448}, with demonstrated wireless data rates exceeding 100 Gbps in controlled environments. Further integration of antenna-in-package (AiP) structures and reconfigurable phased arrays is expected to push link capacities toward the terabit regime. Modern THz RF modules incorporate compact frequency multipliers, low-noise amplifiers, power amplifiers, and high-speed IQ mixers that are specifically optimized for sub-terahertz carrier generation and formed transmission.
    At the device-technology level, THz transceivers rely on a diverse set of electronic and photonic components. CMOS circuits provide low-power digital baseband processing and large-scale integration, while SiGe HBT and III–V HEMT devices deliver the high gain, high frequency, and high linearity required in the RF front end~\cite{11048931}. These electrical-domain devices are increasingly co-designed with photonic components such as optical modulators, photodiodes, and waveguides that support high-speed clock distribution and hybrid electro-optical signal generation, tailored for future data center interconnects.



    \item \textbf{High-speed baseband processing}: Recent advances in baseband architectures enable THz wireless interconnects to achieve both low latency and high energy efficiency. Short and stable data-center links allow the system to move away from heavy DSP pipelines toward streamlined mixed-signal designs. Techniques such as short-symbol OFDM or DFT-s-OFDM, single-carrier modulation, and lightweight equalization significantly reduce processing delay. Hybrid forming and hybrid precoding further lower complexity by shifting most spatial processing to the analog domain, thereby reducing the number of high-power RF chains and lowering ADC/DAC requirements. Low-resolution quantization (e.g., 1–4 bit ADCs) and minimal-iteration FEC further shrink the digital workload while maintaining reliable performance in controlled indoor environments. Collectively, these techniques enable sub-100-ns processing latency and pJ/bit-level energy efficiency, making the baseband subsystem a key enabler for scalable THz data-center interconnects.
    
\end{itemize}


\section{Comparative Analysis: Latency and Energy}
Latency is the defining advantage of THz wireless interconnects. In conventional optical fabrics, packets traverse multiple switch layers, e.g., Top-of-Rack (ToR), aggregation, and core, introducing delays in the microsecond range. Each switch adds serialization, queuing, and optical-electrical conversion delays. By contrast, a THz wireless link provides a direct LoS connection between racks. For a $10$~m link, propagation delay is only about $33$~ns; even with beam alignment and processing overhead, the total latency can stay below $50$~ns.
In terms of energy efficiency, optical transceivers typically consume $10–15$~W per $400$~G module, corresponding to $20–40$~pJ/bit~\cite{6848762}, while copper interconnects with active equalizers can exceed $100$~pJ/bit. By contrast, emerging THz front-ends in short-range environments show simulated energy efficiencies approaching single-digit pJ/bit levels~\cite{9735144}. Within the specific regime of $5–20$~m distances, THz wireless can therefore achieve superior energy-delay products compared to wired alternatives.

\begin{figure}
    \centering
    \subfloat[]{
        \hspace{-0.01\columnwidth}
        \includegraphics[width=0.9\columnwidth]{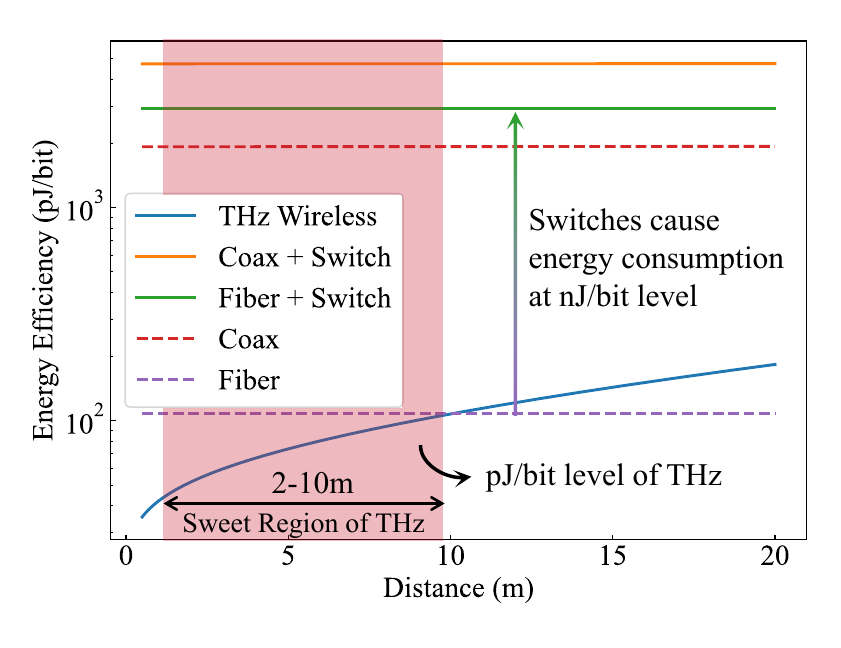}
    }
    \hfill
    \subfloat[]{
        \hspace{0.03\columnwidth}
        \includegraphics[width=0.95\columnwidth]{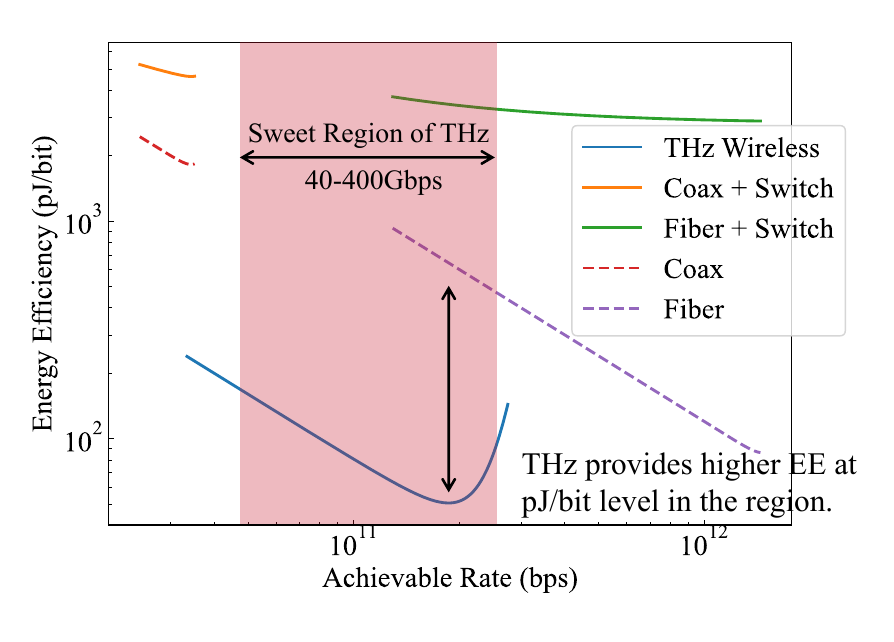}
    }
    \caption{Comparison of energy per bit across copper, optical, and THz regimes.}
    \label{fig:ctfcomp}
\end{figure}

Fig.~\ref{fig:ctfcomp} demonstrates that THz wireless communication can achieve an energy efficiency below 10 pJ/bit within short ranges (e.g., under 20 m), with a gradual increase as distance increases. In contrast, optical fiber maintains nearly constant energy efficiency with distance, but its efficiency drops sharply once switching equipment is introduced, reaching only the order of 1 nJ/bit. 
It is noteworthy that the switch consumes $0.28$~nJ/bit in the model~\cite{6848762}.
In terms of achievable data rates, THz links exhibit a “\textit{sweet region}” at several hundred Gbit/s, where they deliver peak energy efficiency. Although optical fiber can support even higher data rates, its energy efficiency is significantly constrained by the associated switching and transceiver overheads.

Overall, these comparisons highlight the unique positioning of THz wireless as a low-latency, energy-efficient, and high-capacity interconnect solution tailored for next-generation AI clusters. While optical fiber remains unmatched in absolute throughput over long distances, its reliance on multi-stage switching and power-hungry transceivers fundamentally limits its energy efficiency at the rack-to-rack scale. THz links, on the other hand, exploit the short-range, quasi-static environment of data centers to deliver near speed-of-light latency, pJ/bit-level efficiency, and multi-hundred-gigabit data rates with simplified physical-layer design. As device technologies continue advancing, especially in CMOS THz front-ends, reconfigurable antenna arrays, and low-overhead baseband architectures, THz wireless interconnects are poised to evolve from a promising experimental concept into a practical, scalable alternative to traditional wired fabrics for future AI infrastructure.



\section{Challenges and Research Directions}
Despite significant promise, realizing THz wireless data centers entails major challenges. 
First, although THz band operation has been demonstrated across a variety of device platforms, the development of fully integrated and ultra-broadband THz modulator and demodulator circuits is still at an early stage. Present THz front ends face limitations in output power, circuit complexity, and the bandwidth of digital-to-analog converters. These factors restrict the usable intermediate frequency bandwidth and prevent the system from taking full advantage of the very large spectrum available in the THz range. In addition, practical chip-level integration that includes on-chip antennas, advanced packaging, and co-design with optical or electrical interfaces remains a major challenge for building scalable and manufacturable THz transceivers~\cite{nagatsuma2016advances}.


Second, the physical hardware required to sustain nanosecond-level alignment at THz frequencies remains a major barrier. Although angle-of-arrival (AoA) estimation and forming algorithms are well established, the underlying RF hardware—phase shifters, switches, mixers, and LO distribution networks still struggles to operate at the tens-of-gigahertz control rates needed for nanosecond reaction times. THz front ends suffer from limited phase resolution, high insertion loss, and thermal drift, while wideband THz phased arrays require tightly synchronized clocking. Moreover, maintaining stable s in the presence of rack-scale vibrations demands ultra-fast, low-jitter actuation loops that exceed the capabilities of today’s CMOS or SiGe processes. These hardware-driven constraints, rather than algorithmic limitations, now form the dominant bottleneck for reliable, reconfigurable THz connectivity inside future data centers.


Third, baseband architectures must deliver terabit-per-second processing with extreme energy efficiency while remaining tightly integrated with existing fiber interconnects. Achieving sub-10 pJ/bit operation at THz bandwidths requires radical departures from conventional DSP pipelines: converters must support tens of gigahertz of instantaneous bandwidth, memory interfaces must sustain multi-Tb/s data flow, and mixed-signal baseband designs must balance quantization noise with power consumption. These requirements push the limits of ADC/DAC technology, mixed-signal clocking, and on-chip interconnect fabrics. At the system level, future data centers will need unified hardware that merges high-capacity optical backbones with flexible THz wireless fabric, allowing tasks, data paths, and s to be scheduled in real time across both media. Building such a deeply integrated, fiber-compatible, and dynamically orchestrated wireless-optical architecture remains one of the core challenges on the road to practical THz-enabled web-scale clusters.

Furthermore, accurate modeling of THz propagation in metallic rack environments with reflections, scattering, and blockage is crucial to predict performance. 
Data centers form a unique electromagnetic environment dominated by rows of metallic server racks, cable trays, perforated panels, and active cooling airflow. These structures generate complex multipath components, angle-dependent reflections, and frequency-selective fading, while even a passing operator or robotic unit can temporarily block the LoS path. According to our preliminary measurement, the OLoS case can reach over 50\,\%. 
Accurate system design requires high-fidelity deterministic modeling (e.g., ray-tracing with precise point clouds and material parameters), dynamic blockage analysis, and integration with DT environments that capture changes in rack layout or cooling conditions. Such modeling capabilities are essential for link budgeting, topology optimization, and robust beam management strategies.

Finally, real-time orchestration across thousands of dynamic wireless links will require more than traditional software-defined networking. It demands an AI-driven control fabric tightly coupled with a digital DT that continuously mirrors the physical data-center environment. As THz links, and eventually quantum-enhanced THz channels, are reconfigured in response to workload movement and shifting traffic patterns, the DT provides a live, predictive view of channel conditions,  alignment stability, and impending link degradation. This virtual replica allows AI controllers to evaluate reconfiguration strategies before they are executed, ensuring stability while managing thousands of s and routing paths. By integrating these predictions with the data-center scheduler, task placement and network topology can co-evolve in real time, enabling sub-millisecond, cross-layer adaptation. Such AI-DT coordination will be essential for future data centers as they transition toward large-scale THz and, ultimately, THz-quantum-wireless fabrics capable of supporting highly dynamic, AI-driven computing workloads.

Moving forward, several research directions emerge. Demonstration of full-stack THz data-center prototypes is a critical milestone, targeting \textgreater\,400 Gbps throughput over \textgreater\,10 m links at \textless\,100 ns latency. Integration of DTs with THz hardware can enable self-optimizing networks that adapt beam patterns and frequencies in real time. Hybrid optical-THz fabrics represent another key direction: optical fibers serve as the static backbone, while THz links act as the dynamic overlay for workload hotspots and reconfigurable pods. Finally, international standardization (IEEE 802.15.3d, ITU-R 6G-THz task groups) and joint academia–industry consortia will be essential for interoperability and spectrum harmonization.
Additionally, an exciting direction lies in extending THz wireless technologies toward emerging quantum data center architectures, where ultra-high-frequency, low-noise, and reconfigurable THz links could enable short-range quantum state distribution, hybrid quantum–classical interconnects, and wireless quantum control planes. Such developments may ultimately lead to the first generation of quantum-enabled wireless data centers, pushing the limits of both computation and communication.

\section{Conclusion}

THz wireless interconnects have the potential to fundamentally reshape AI data-center architectures by delivering one-hop connectivity, ultra-low-latency, and unprecedented energy efficiency. Although today’s implementations still fall short of the long-term vision, rapid progress across THz device technologies, materials, and baseband/forming architectures, including low-power RF front-ends, high-speed hybrid precoding, and measurement-driven channel modeling, is steadily advancing toward this vision. By tightly integrating these wireless capabilities with the massive throughput of optical fabrics, future data centers can achieve both scalable performance and sustainable operation. In parallel, the industry’s transition toward modular compute creates an additional opportunity where THz links serve as a natural, non-intrusive medium for interconnecting and validating these components. Together, these advances signal the emergence of a wireless-defined compute era, where agility, low latency, and energy efficiency become foundational to next-generation AI infrastructure.

\bibliographystyle{IEEEtran}
\bibliography{DC}

\newpage

 




\vfill

\end{document}